\documentclass[12pt]{article}
\usepackage[super]{natbib}
\usepackage{graphicx}

\begin{document}

\title{Statistical uncertainty in the re-analysis of polarization in GRB021206}

\author{Steven E. Boggs$^{\dag,*}$ \& Wayne Coburn$^{\dag}$}

\date{Submitted to MNRAS, 17 October 2003}

\maketitle

{\small\noindent$^{*}$Department of Physics, and\\
$^{\dag}$Space Sciences Laboratory, University of California at Berkeley,\\
Berkeley, CA, 94720, USA}

\vspace*{0.5in}

%%\author{Wayne Coburn\altaffilmark{1} \&
%%Steven E. Boggs\altaffilmark{1}$^,$\altaffilmark{2}}

%%\altaffiltext{1}{Space Sciences Laboratory,
%%University of California at Berkeley, Berkeley, CA, 94702-7450, USA}

%%\altaffiltext{2}{Department of Phyisics,
%%University of California at Berkeley, Berkeley, CA, 94702-7450, USA}

%%\received{2003 February 7}
%%\accepted{2003 March 12}

\noindent{\bf Abstract}

We recently reported the first detection of an astrophysical
gamma-ray polarization
from GRB021206 using the \textit{Reuven Ramaty} High Energy Solar Spectroscopic
Imager (RHESSI) spacecraft. Our analysis suggested gamma-ray polarization at
an astoundingly high level, 80$\pm$20\%.
A recent manuscript re-analyzes this event in the RHESSI data, and sets an upper limit on
potential polarization of 4.1\% -- clearly inconsistent with our
initial analysis. This manuscript
raises a number of important concerns about the analysis, which are already
being addressed in a separate methods paper under preparation. We note
here, however, that the limit set on the
potential polarization by this re-analysis is significantly underestimated using
their novel statistical methods.

\noindent{\bf Introduction}

GRB021206 was detected by both RHESSI and
the Interplanetary Network (IPN) on 6 December 2002
at 22:49 UT \citep{b2,b3}. The GRB was immediately identified as both exceptionally
bright and hard. The IPN localized the GRB to 18$^{\circ}$ off solar.
The combination of a bright hard burst and proximity to the RHESSI roll axis made
this GRB an ideal candidate to search for polarization with RHESSI
using detector-detector coincidence events. The idea behind this analysis
is straight forward \cite[CB03 hereafter]{b1}. Polarized gamma-rays will preferentially
scatter at right angles relative to the direction of their polarization.
Therefore, by looking for a preferential scatter axis in the detector
coincidence data (some fraction of which are photons which
have truly scattered from one detector to another) we can search for signs
of an intrinsic polarization in the gamma-rays. Our initial analysis
of this event suggested an intrinsic polarization of 80 $\pm$ 20\% (CB03).

In practice this analysis is not as easy as outlined above due to
two complicating factors.
First, RHESSI is not designed to tag coincidence events.
Therefore, detector coincidences must be reconstructed from the
individual event data by comparing time tags on each event and defining
a coincidence time window. In addition, for large count rates like those
during GRB021206, a significant fraction of these coincidence events
are accidental coincidences, and not true scatter events. Therefore, these
relative rates must be carefully considered in any analysis designed to
study polarization.

A recent re-analysis of the RHESSI data has found a significantly
different number of detector coincidence events than the number presented in
our initial paper \cite[RF03 hereafter]{b4}.
Given the sensitivity of the derived detector-coincidence rate to both
the data cuts and logic, it is not surprising that a smaller number of
coincidence events, and specifically a dramatically smaller number of
inferred true scatter events, were derived in RF03. Our original paper
did not allow a detailed description of our event cuts and logic; however,
we are currently preparing a methods paper which will cover these cuts
in greater detail.

We are concerned, however, that RF03 go on to place a very tight upper limit
on the potential polarization of GRB021206 of $\le 4.1\%$ (90\% confidence). Using their
derived numbers on coincidence events, this limit
is in clear contradiction with the most basic counting statistics.
If we define $S$ as the total number of true detector-detector scatter events,
and $B$ as the total number of background coincidence events (both chance
coincidences and true background photon scatters), then the expected signal-to-noise
ratio, $\sigma$, for measuring a fractional polarization $\Pi$ is given by
the simple formula \citep{b5,b6,b7},
\begin{equation}
\sigma = \frac{\mu \Pi S}{\sqrt{2(S+B)}} ,
\end{equation}
where $\mu$ is the instrumental modulation factor which can be thought of
as the effective fraction of scattered, polarized photons which contribute to
a measured modulation. This formula is simple to understand in terms of basic counting
statistics. The
numerator ($\mu \Pi S$) is just the total number of polarization signal counts (i.e. counts
potentially contributing to a measurable modulation), and the denominator ($\sqrt{2(S+B)}$)
is the square root of twice the total number of counts,
source + background, which is just the
noise level on the overall measurement  (some derivations drop the $\sqrt{2}$ factor).
The modulation factor $\mu$ was clearly identified
in CB03 as the largest systematic uncertainty on the measurement. By definition, $\mu \le 1$, and
for most real gamma-ray instruments it is much less than unity. For RHESSI we estimated
that $\mu = 0.19 \pm 0.04$.

We can turn this formula around to determine the minimum detectable
polarization (or an upper limit) given the measured number of counts $S$ and $B$.
For a 90\% confidence level (used in RF03),
we must set the upper limit a factor of 1.65$\sigma$ above the noise level. Therefore, the
minimally detectable polarization, or 90\% confidence upper limit on a null polarization measurement,
is given by:
\begin{equation}
\Pi = 1.65 \frac{\sqrt{2(S+B)}}{\mu S}.
\end{equation}
Using the numbers from RF03, they deduced that $S+B = 8230$ counts and $S = 830 \pm 150$
counts. Plugging these numbers in the formula above, and using our estimate of the modulation
factor $\mu$, yields an upper limit on their measured polarization of $\Pi < 130\%$ (90\%
confidence). This simple estimate is in stark contrast with their stated upper limit
of 4.1\%. In fact, if we replace ($\sqrt{2(S+B)}$) by their stated uncertainty on $S$, then
we derive an even higher upper limit of $\Pi < 160\%$ (90\% confidence).
By contrast, in our analysis we found $S+B = 14916$ counts, and $S = 9840 \pm 96$
counts (CB03), which leads to a minimum detectable polarization of 15\% (not including
systematics).

\section{Conclusion}
While we are intrigued by the novel statistical technique presented in RF03 (and
especially that the inferred polarization level is independent of instrumental
response), we are led to the inevitable conclusion that there is a serious flaw in their
statistical method. At most, RF03 can claim that their analysis is insensitive to polarization
at any level, and therefore not inconsistent with the level of polarization presented in
our original paper.

\vspace*{1em}
\noindent{\bf Correspondence} and requests for materials should be addressed
to S.E.B. (boggs@ssl.berkeley.edu) or W.C. (wcoburn@ssl.berkeley.edu)

\bibliographystyle{unsrt}
\bibliography{mnras}

\end{document}